\documentclass[useAMS,usenatbib]{mn2e}
\usepackage{color}
\usepackage{graphicx}
\usepackage{journals}
\usepackage[hyperindex]{hyperref}
\usepackage{arydshln}
 \usepackage{epsfig}
\usepackage{amsmath}
\usepackage{url}
\usepackage{fixltx2e}
\newcommand{\hompc}{\,h\,{\rm Mpc}^{-1}}
\newcommand{\mpcoh}{\,h^{-1}\,{\rm Mpc}}

\newcommand{\simgt}%
{\,\hbox{\lower0.6ex\hbox{$\sim$}\llap{\raise0.6ex\hbox{$>$}}}\,}
\newcommand{\simlt}%
{\,\hbox{\lower0.6ex\hbox{$\sim$}\llap{\raise0.6ex\hbox{$<$}}}\,}
\newcommand{\correction}{}
\pdfoutput=1

\title{Reconstruction in Fourier space}
\author[A.Burden et al.]{
\parbox{\textwidth}{
A. Burden$^{1}$\thanks{E-mail: angela.burden@port.ac.uk}, 
W. J. Percival$^{1}$, C. Howlett$^{1}$}\\
\vspace*{2pt} \\
$^{1}$Institute of Cosmology and Gravitation,  University of Portsmouth, Dennis Sciama building, Portsmouth, PO1 3FX}

\begin{document}

\date{}

\pagerange{\pageref{firstpage}--\pageref{lastpage}} \pubyear{}

\maketitle

\label{firstpage}

\begin{abstract}
We present a fast iterative FFT-based reconstruction algorithm that allows for nonparallel redshift-space distortions (RSD). We test our algorithm on both N-body dark matter simulations and mock distributions of galaxies designed to replicate galaxy survey conditions. We compare solenoidal and irrotational components of the redshift distortion and show that an approximation of this distortion leads to a better estimate of the real-space potential (and therefore faster convergence) than ignoring the RSD when estimating the displacement field.
Our iterative reconstruction scheme converges in two iterations for the mock samples corresponding to BOSS CMASS DR11 when we start with an approximation of the RSD.
The scheme takes six iterations when the initial estimate, measured from the redshift-space overdensity, has no RSD correction. Slower convergence would be expected for surveys covering a larger angle on the sky.
We show that this FFT based method provides a better estimate of the real space displacement field than a configuration space method that uses finite difference routines to compute the potential for the same grid resolution. Finally we show that a lognormal transform of the overdensity, used as a proxy for the linear overdensity, is beneficial in estimating the full displacement field from a dense sample of tracers. However the lognormal transform of the overdensity does not perform well when estimating the displacements from sparser simulations with a more realistic galaxy density.

\end{abstract}
\begin{keywords}
large-scale structure of Universe -- distance scale
\end{keywords}
\section{Introduction}

Numerous upcoming galaxy surveys including WEAVE \citep{WEAVE}, 4MOST \citep{4MOST}, DESI \citep{DESI}, eBOSS (Dawson et. al. in prep.), EUCLID \citep{EUCLID} and WFIRST \citep{WFIRST} will measure the baryon acoustic feature to high precision at a range of redshifts. Their goal is to track the expansion history of the Universe to sufficient precision to rigorously test cosmological models and decipher the mechanism behind the late-time accelerated expansion rate of the Universe. The Baryon Acoustic Oscillation (BAO) scale provides a standard ruler in the distribution of mass and in turn galaxies, providing a tool to make such measurements.

The BAO feature is created when sound waves propagating through the photon-baryon fluid stall as the photon pressure is released during recombination (e.g. \citealt{Meiksin99}).
The stalled sound wave deposits baryonic perturbations in spherical overdense shells around dark matter perturbations. This generates a preferred scale in the matter overdensity distribution, the baryon acoustic feature, large enough ($\approx120\mpcoh$) to remain detectable at low redshifts in the distribution of a large-volume sample of galaxies. Statistically the feature emerges in the 2-point correlation function as a bump or in the power spectrum as a series of peaks and troughs in the amplitude. 

Although the signal is robust, bulk flow motions and non-linear structure formation statistically blur the boundaries of the shell increasingly to lower redshifts \citep{Eisenstein07}. The process of reconstruction recovers some of the lost precision due to these effects and is therefore crucial to fully exploit future galaxy survey data.

The simple reconstruction method of \cite{Eisenstein:2006nk} has been used in numerous analyses \citep{RossMGS,Anderson14_DR11,Tojerio14,Ross14_redblue, Anderson14_DR9,Kazin14,Padmanabhan12} resulting in tighter distance measurements from the baryon acoustic signal. The method relies on the ability to recover the displacements of overdensities from their initial positions and reverse this motion. Practically this is achieved by moving galaxies and a set of random particles that trace the geometry of the survey back along these vectors. Statistically this sharpens the measured signal and has been shown to remove a slight bias in the distance measurement induced by non-linear structure formation \citep{Eisenstein07, Mehta11}.

It is straightforward to compute an estimate of the displacement field from the smoothed real space overdensity using the Zel'dovich approximation \citep{Zeldovich} and Fast Fourier Transforms (FFTs). However, complications arise due to RSD which distort the measured overdensity from the true overdensity. 
One approach to include RSD is to use finite difference methods to linearly model both real space potential and RSD \citep{Padmanabhan12}. The real space potential is recovered at each location from the overdensity field measured in redshift-space using a linear equation solver.
The displacement field is computed from the potential using finite differences. The accuracy of this method is limited by the grid resolution adopted. Furthermore, solving for the potential on the grid using a linear equation solver is computationally expensive.

Although one cannot directly compute the real space displacement field from the redshift-space overdensity using FFTs because of the varying line of sight, \cite{Burden14} made an approximation that allows partial correction of the redshift-space component. The resulting displacement vectors were shown to be well matched to those computed in configuration space. We now show that we can improve upon this by developing an iterative scheme to correct for RSD.

Both FFT and finite difference methods require us to use the overdensity field to estimate the displacements. {\correction \cite{Neyrinck09} show that the lognormal transform of the overdensity field measured in N-body simulations reduces the non-linear component of the field on quasi-linear scales thereby recovering more linear information in the power spectrum. The lognormal transform of the field is more Gaussian, thus off diagonal elements of the covariance matrix are reduced. In \cite{Seo11} the authors apply the lognormal transform to the weak-lensing convergence field and find similar results. The nonlinear information in the convergence field is reduced and the information content in the covariance matrix enhanced.} Furthermore it has been shown \citep{Falck12} that the divergence of the true displacement field (using N-body simulations) is better replicated under certain conditions by a logarithmic transform of the overdensity. We now test whether this improvement also helps with reconstruction.

In Section~\ref{sec:reconmethods}, we outline two basic methods used to compute the displacement vectors in the reconstruction process. 
In Section~\ref{sec:data} we describe the data sets used to test our algorithms. 
In Section~\ref{sec:effectsirrot}, we test how well the the displacements field computed from the methods in Section~\ref{sec:reconmethods} match the real space displacement field. In Section~\ref{sec:iterative}, we present our iterative reconstruction algorithm. In Section~\ref{sec:lognormal}, we test the effect of using a lognormal transform of the overdensity to compute the displacement vectors in real space. 
Finally in Section~\ref{sec:conclusion} we conclude by combining the results of our tests to offer our most effective method of recovering the real space first order Lagrangian displacement field from an evolved galaxy overdensity field measured in redshift-space.

\section{Reconstruction Methods}\label{sec:reconmethods}

The reconstruction process requires computing the Lagrangian displacement field from the evolved galaxy overdensity distribution measured in redshift-space. In this section we outline two methods to do this.

A Lagrangian co-ordinate system tracks the motion of a fluid element through space and time where the Lagrangian displacement vector ${\bf \Psi}({\bf q}, t)$ maps a fluid at some initial position ${\bf q}$ to its Eulerian position some time later ${\bf x}({\bf q},t)$,
\begin{equation}\label{eq:Lagrange}
{\bf x}({\bf q},t) = {\bf q} + {\bf \Psi}({\bf q}, t).
\end{equation}
The overdensity field is defined as $\delta({\bf x}) \equiv (\rho({\bf x}) - \bar{\rho})/ \bar{\rho}$ and relates the density, $\rho$, at location ${\bf x}$ to the expected average density, $\bar{\rho}$.

The Zel'dovich approximation \citep{Zeldovich} is the first order term in a perturbative expansion of ${\bf \Psi}({\bf q}, t)$ in Lagrangian perturbation theory \citep{Zeldovich,Buchert93,Buchert94,Bouchet94,Catelan95,Taylor96},
\begin{equation}\label{eq:Zeldovich}
{\bf \Psi}({\bf q})^{(1)} = \int \frac{d^3k}{(2\pi)^3}\frac{ i{\bf k}}{k^2} \delta^{(1)}({\bf k}) e^{i{\bf k \cdot q}}.
\end{equation}
The successive perturbative terms ${\bf \Psi^{(n)}}$ are functions of powers of the linear overdenisty perturbation.
We use the Zel'dovich approximation throughout this work, and therefore drop the superscript.

To extract the displacement vector field from a redshift-space galaxy overdensity field, one must solve the equation \citep{Nusser94}
\begin{equation}\label{eq:Nusser}
\nabla \cdot {\bf \Psi} +  f\nabla \cdot ({\bf \Psi\cdot \hat{r}} ){\bf \hat{r}} = \frac{-\delta}{b}.
\end{equation}
This can be solved on a grid in configuration space, computing gradients using finite difference approximations. However, solving for a value of ${\bf \Psi}$ at each grid point is computationally expensive and the accuracy of the method is dependent on the resolution of the grid. Alternatively, \cite{Burden14} made the assumption that  $ ({\bf \Psi\cdot \hat{r}} ){\bf \hat{r}} $ is an irrotational field and use FFTs to compute the value of ${\bf \Psi}$ increasing the speed of the computation. We now outline these two methods.

\subsection{Finite difference approximations}\label{sub:FD}
The configuration space method solves for the potential, $\phi$. Assuming ${\bf \Psi}$ is an irrotational field, as expected on linear scales, this is related to the first order Lagrangian displacement as ${\bf \Psi} = -\nabla \phi$ and Eq.~\ref{eq:Nusser} can be written as
\begin{equation}\label{eq:mainpot}
\nabla^2 \phi + f \nabla\cdot \left(\nabla\phi_r\right) \mathbf{\hat{r}} = \frac{-\delta_g}{b}.
\end{equation}
We follow \cite{Padmanabhan12} and solve this on a grid using finite differences to approximate the derivatives.
The potential at each grid point is expressed as a function of the potential at the surrounding grid points. The Laplacian of the potential at a grid point can be approximated as a function of the potential at the 6 nearest grid points and the central point
\begin{equation}
\nabla^2 \phi _{000} \approx \frac{1}{g^2} \left[ \sum_A \phi_{ijk}  -6\phi_{000}\right],
\end{equation}
where the sum over $A$ is the sum over the 6 adjacent grid points and $g$ is the spacing between grid points.
The second part of Eq.~\ref{eq:mainpot} can be written as
\begin{equation}
f \nabla\cdot \left(\nabla\phi_r\right) \mathbf{\hat{r}} = f\left(\mathbf{ \hat{r} \cdot \nabla \left(\nabla \phi_r \right)} + \mathbf{ \nabla} \phi_r \left(\mathbf{\nabla \cdot \hat{r}}\right) \right),
\end{equation}
which can be approximated as 
\begin{equation}
-2\frac{\phi_{000}}{g^2} + \sum_B f\left(\frac{x_i^2}{g^2r^2} \pm  \frac{x_i}{g r^2}\right) \phi_{A} + \sum_C (-1)^p f \frac{x_i x_j }{2 g^2 r^2}\phi_B,
\end{equation}
where $B$ is the set of points $ijk$ such that 2 of the indices are zero and the other is $\pm 1$, $x_i$ is the cartesian position of the non-zero index and $r$ is the distance to the central grid point. $C$ is the set of points where two of the indices are $\pm 1$ and the third is zero. When the two indices are the same, $p=0$, when they are different $p=1$. $x_i$ and $x_j$ are the cartesian positions of the non-zero indices. 

This can be arranged as a linear system of equations such that $\mathbf{A \phi} = \delta_g/b$, where $\mathbf{A}$ is a matrix describing the dependence of the potential on its surroundings. The $\delta$ that we input here is the smoothed overdensity field measured in redshift-space. We solve for the potential using the GMRES in the PETSc package \citep{petsc-web-page,petsc-user-ref}. Once we have a model for the potential, finite differences are used again to calculate the displacements at each grid point from the potential.

\subsection{Fast Fourier Transform method with redshift-space approximation}

While ${\bf \Psi}$ is expected to be irrotational, $({\bf \Psi} \cdot {\bf \hat{r}}){\bf \hat{r}}$ is not. This means we cannot easily solve Eq.~\ref{eq:Nusser} using Fourier methods. To see this we decompose $({\bf \Psi} \cdot {\bf \hat{r}}){\bf \hat{r}}$, using Helmholtz's Theorem, into the gradient of a scalar potential field and the curl of a vector field 
\begin{equation} \label{eq:Helmholtz}
\left(\mathbf{\Psi\cdot \hat{r}}\right)\mathbf{\hat{r}} = \nabla A + \nabla \times {\bf B},
\end{equation}
where A is a scalar potential field and and ${\bf B}$ is a vector potential field. We refer to $\nabla A$ as the irrotational component and $\nabla \times {\bf B}$ as the soleniodal component. Substituting Eq.~\ref{eq:Helmholtz} into Eq.~\ref{eq:Nusser} and writing ${\bf \Psi}$ as the gradient of a potential as in Eq.~\ref{eq:mainpot} we get
\begin{equation}\label{eq:approx1b}
\nabla (\phi + fA) = -\nabla \nabla^{-2} \frac{\delta_g}{b}.
\end{equation}
This is exact and without approximation, but does not allow us to calculate ${\bf \Psi}$ {\correction directly using FFTs}. A further assumption.
\cite{Burden14} made the approximation that $ ({\bf \Psi\cdot \hat{r}} ){\bf \hat{r}} $ is irrotational, that is
\begin{equation}\label{eq:irrot}
 \left(\mathbf{\Psi\cdot \hat{r}}\right)\mathbf{\hat{r}} \approx \nabla A,
\end{equation}
making Eq.~\ref{eq:approx1b} 
\begin{equation}\label{eq:approxpsi}
{\bf \Psi} + f ({\bf \Psi} \cdot {\bf \hat{r}}) {\bf \hat{r}} = -\nabla \nabla^{-2} \frac{\delta_g}{b}.
\end{equation}
The RHS can be computed in Fourier space using FFTs, and the solution to this equation is
\begin{equation}\label{eq:approx1}
{\bf \Psi } = {\rm IFFT} \left[- \frac{i\mathbf{k }\delta\left(\mathbf{k}\right)}{k^2 b}\right] - \frac{f}{1+f} \left\{ {\rm IFFT} \left[- \frac{i\mathbf{ k }\delta\left(\mathbf{k}\right)}{k^2 b}\right] \cdot{\bf \hat{r}} \right\} {\bf \hat{r}},
\end{equation}
where IFFT is the inverse Fourier Transform.

In Section~\ref{sec:effectsirrot} we show that this approximation can be improved by accounting for the solenoidal component. In Section 5 we show that the approximations do not change the outcome of our new iterative process designed to solve Eq.~\ref{eq:Nusser} numerically, although they do speed up convergence.

\subsection{Displacement field nomenclature}\label{sub:list}
As we use a number of estimates of the displacement field computed using a variety of methods, we introduce our notation here;
\begin{itemize}
\item ${\bf \Psi}_{\textrm{true}}$, the true displacement field measured from final minus initial particle positions (N-body only).
\item ${\bf \Psi}_{\textrm{red, FD}}$ displacements computed from the redshift-space overdensity in configuration space using the finite difference (FD) method.
\item ${\bf \Psi}_{\textrm{real, FD}}$ displacements computed from the real space overdensity in configuration space.
\item ${\bf \Psi}_{\textrm{real, FFT}}$ displacements computed using FFTs and the real space overdensity.
\item ${\bf \Psi}_{\textrm{red, FFT}}$ displacements computed using FFTs and the redshift-space overdensity (with no RSD correction).
\item ${\bf \Psi}_{\textrm{FFTA}}$ displacements computed using FFTs and the redshift-space overdensity where an approximate correction assuming that $({\bf\Psi}\cdot{\bf\hat{r}}){\bf \hat{r}}$ is irrotational has been applied.
\end{itemize}

 \begin{figure*}
     \centering
    \resizebox{0.95\textwidth}{!}{\includegraphics{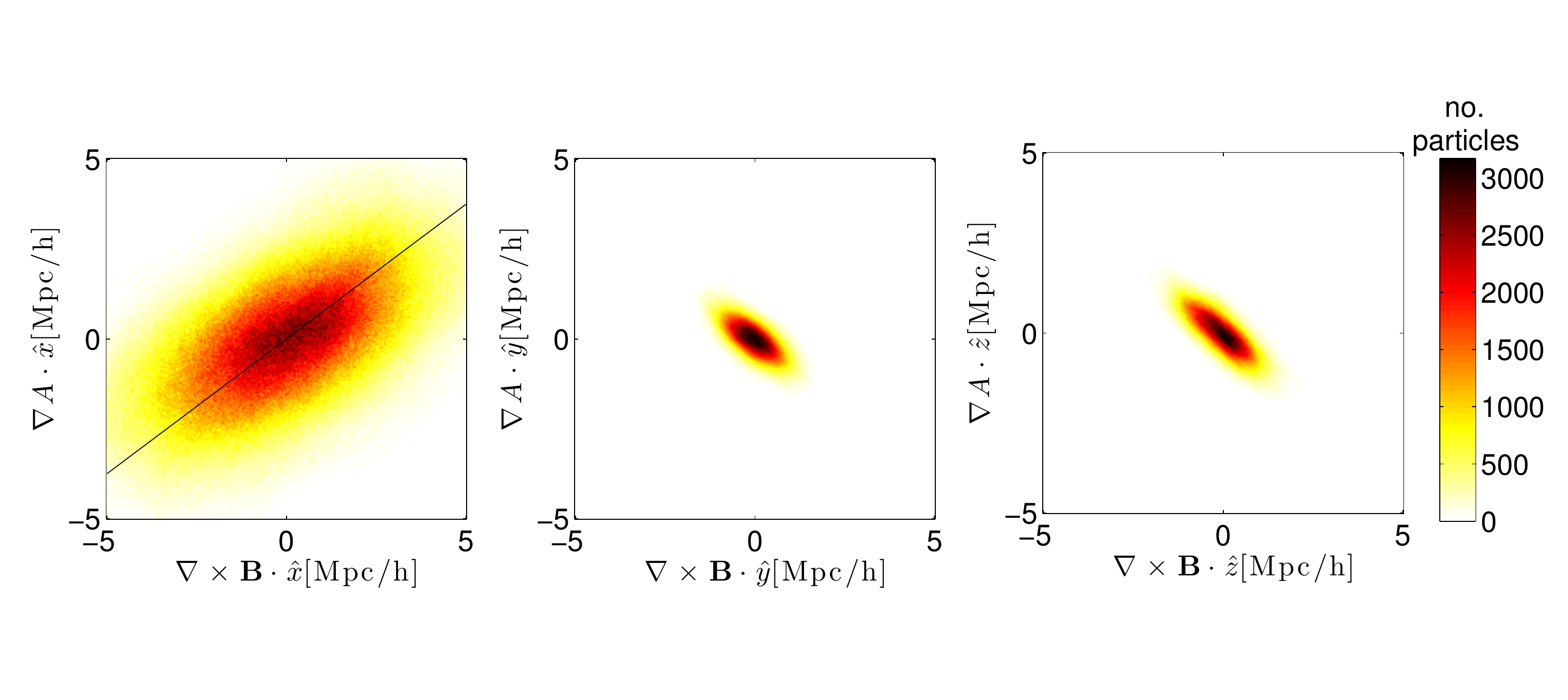}}
 \caption{The plots show a comparison of irrotational and solenoidal components of $({\bf \Psi} \cdot {\bf \hat{x}}){\bf \hat{x}}$ projected in cartesian directions. {\correction The redshift space distortions are modelled as plane-parallel and projected along the ${\bf \hat{x}}$ direction.} The components projected along ${\bf \hat{x}}$ (left) are positively correlated and the amplitudes are higher than in the other two directions. The average ratio of these vectors is 0.75, this is shown by the black line. The sum of the two components projected in the ${\bf\hat{y}}$ (centre) and ${\bf \hat{z}}$ (left) is expected to be zero, the components are negatively correlated in these two directions. The amplitudes of these components is small. }
 \label{fig:Helmholtz_hist}
\end{figure*}
\section{Catalogues}\label{sec:data} 
In this section we describe the two sets of simulations used to test our algorithms. 
\subsection{N-body simulations}
We use a TreePM N-body simulation box of $1024^3$ dark matter particles generated with GADGET-2 \citep{Springel05,Springel01} with initial positions at $z=49$ and final positions and velocities at $z=0$. The box has side length $768\mpcoh$ comoving coordinates. 
The fiducial cosmology of the simulation has $\Omega_m =0.317$, $h=0.670$, $\Omega_bh^2 = 0.022$.

The full sample is randomly subsampled to $20(128^3)$, $5(128^3)$, $128^3$ and $0.25(128^3)$ particles to create simulations with varying density. {\correction All of these samples are derived from a single realisation}. For brevity these are named the 20t, 5t, 1t and 0p25t samples respectively. 
In order to investigate redshift-space effects we locate an observer at $z=0$, and modify the radial positions by $({\bf v \cdot \hat{r}})/\mathrm{H}_0$ where ${\bf v}$ is the final velocity in units $\mathrm{kms}^{-1}$ and the Hubble constant is $\textrm{H}_0$ is 100 $\mathrm{kms}^{-1} \hompc$ {\correction and ${\bf \hat{r}}$ is the unit vector pointing away from the origin of the box.} This gives a sample with strong wide-angle effects, a very wide sky area (1/8 of the sky) near to the observer.

The particle overdensity is measured as described in Section~\ref{sec:reconmethods} where $\rho({\bf x})$ is the density of the particle distribution at grid point ${\bf x}$ and $\bar{\rho}$ is the average particle density per grid point. We use a NGP binning scheme on a $128^3$ mesh. {\correction We do not investigate alternative binning methods as the grid points are smoothed post binning mitigating any effects of different binning schemes.}

Gaussian smoothing filters are applied in Fourier space as a multiplicative term $S(k) = \exp {-ik^2R^2/2}$ where $R$ is the smoothing length. Smoothing removes non-linearities in the overdensity, reduces shot noise and sets the scale of the overdensities to be moved during the reconstruction process. The smoothed overdensity is used as a proxy for the linear overdensity to calculate the first order Lagrangian displacement field in Eq.~\ref{eq:Zeldovich}.

\subsection{Mock survey catalogues}
To conduct our tests on galaxies with realistic masks, we make use of the publicly available PTHalo \citep{Manera13} mock samples created to match Data Release 11 (DR11) of the Baryon Oscillation Spectroscopic Survey (BOSS) galaxy samples \citep{Dawson13}. 

We use the CMASS Northern sample spanning a redshift range of $0.43\leq z\leq0.70$ and covering an effective area of 6,308 square degrees.
The PTHalo method initially creates a matter field based on second order Lagrangian Perturbation Theory (LPT). Halos are located with a Friends of Friends (FoF) algorithm, and halo masses calibrated to N-body simulations. The halos are populated with galaxies using a Halo Occupation Distribution (HOD) calibrated by the observed galaxy samples on small scales between $30\mpcoh$ and $80\mpcoh$ {\correction using the 2-point statistics of the true galaxy sample.} Redshift-space distortions are added to the mock galaxy distribution by modifying their redshifts according to the second order LPT peculiar velocity field in the radial direction. The matter field is created in a single time-slice, rather than in a light cone, thus the growth rate and RSD signal are constant throughout the sample. The effective redshift of the survey is $z=0.57$. 

The mock catalogues are sampled to match the angular mask and redshift cuts of the survey data. To replicate observational complications, galaxies are subsampled to mimic missing galaxies caused by redshift failure, and close pairs as simultaneous spectroscopic observations are limited to objects separated by greater than $62^{\prime\prime}$. 

The fiducial cosmology of the mock sample has $\Omega_m=0.274$, $h=0.70$, $\Omega_bh^2 =0.022$, $n_s = 0.95$ and $\sigma_8 =0.80$.
We assume a local deterministic galaxy bias such that $\delta_g = b\delta$, where $b=1.87$ is the galaxy bias estimated empirically from the real survey data. 

As the distribution of mock galaxies includes sampling and mask effects of survey data we compute the overdensity using a random catalogue of particles Poisson sampled within the survey mask. To minimise shot noise the random catalogue contains 100 times more data points than the mock catalogue.

{\correction Although we do not compute the power spectrum in this work,} we weight mock galaxies using the FKP weighting scheme in \cite{FKP}, designed to optimally recover the overdensity field given a varying density sample. We therefore apply a weight to each galaxy
\begin{equation}
  w = w_{FKP} \left(w_{cp} + w_{red} -1\right),
\end{equation}
where $w_{cp}$ and $w_{red}$ correct for the close-pairs and redshift failures respectively (see \citealt{Anderson14_DR9} for further details), and $w_{FKP}$ is the FKP weight
\begin{equation}
w_{FKP} = \frac{1}{1 + \bar{n}\left(z\right)P_{0}},
\end{equation}
with fixed expected power spectrum $P_0 = 20,000h^{-3}\mathrm{Mpc}^3$, and average galaxy density $\bar{n}\left(z\right)$.

The galaxies are set in a box of length $3500\mpcoh$ and the overdensity is computed using the NGP scheme on a $512^3$ mesh. The mock catalogue is sparse, therefore we smooth the galaxy and random catalogue density prior to computing the overdensity. To prevent spurious spikes in our overdensity post smoothing, we multiply the smoothed overdensity by a binary mask that sets all grid points that fall outside of the survey region to zero. 
The first order displacement field is computed as before.

\section{The effects of an irrotational approximation}\label{sec:effectsirrot}

In this section we investigate the RSD term in Eq.~\ref{eq:Nusser}, $({\bf \Psi} \cdot {\bf \hat{r}}){\bf \hat{r}}$, and in particular consider the approximation that it is irrotational.
We show that this approximation can be improved beyond that of \cite{Burden14}.

\subsection{The plane-parallel framework}
Although ${\bf \Psi}$ is irrotational \citep{Bouchet95}, $({\bf \Psi \cdot \hat{r}}) {\bf \hat{r}}$ is not and the true vector field has both irrotational and solenoidal components as shown in Eq.~\ref{eq:Helmholtz}.
To estimate the amplitude of these components we assume a distance observer plane parallel approximation of the RSD where ${\bf \hat{r}} \rightarrow {\bf \hat{x}} $. Taking the curl of $ ({\bf \Psi\cdot \hat{x}} ){\bf \hat{x}}$ gives
\begin{equation}
\nabla \times ({\bf \Psi\cdot \hat{x}} ){\bf \hat{x}} = \nabla \times (\nabla \times {\bf B}).
\end{equation}
Therefore in Fourier space the solenoidal components of the redshift-space distortion of the displacement field are
\begin{equation}\label{eq:solenoidal}
\nabla \times {\bf B} =  \frac{1}{k^2} \left( k_y^2 + k_z^2\right) \Psi_x {\bf \hat{k}_x}  - \frac{k_y k_x }{k^2} \Psi_x {\bf \hat{k}_y} - \frac{k_z k_x }{k^2} \Psi_x {\bf \hat{k}_z},
\end{equation}
and the irrotational components are
\begin{equation}\label{eq:irrotational}
\nabla A =  \frac{1}{k^2} k_x^2\Psi_x {\bf \hat{k}_x}  + \frac{k_y k_x }{k^2} \Psi_x {\bf \hat{k}_y} +\frac{k_z k_x }{k^2} \Psi_x {\bf \hat{k}_z},
\end{equation}
where we have made the abbreviation $\Psi_x ={\bf \Psi\cdot \hat{x}}$.
We see that $\nabla \times {\bf B}$ is non-zero and has components of the same order of magnitude as $\nabla A$. As expected, off-axis components cancel in the sum.

\begin{figure}
 \hspace{-0.5cm}
    \resizebox{0.5\textwidth}{!}{\includegraphics{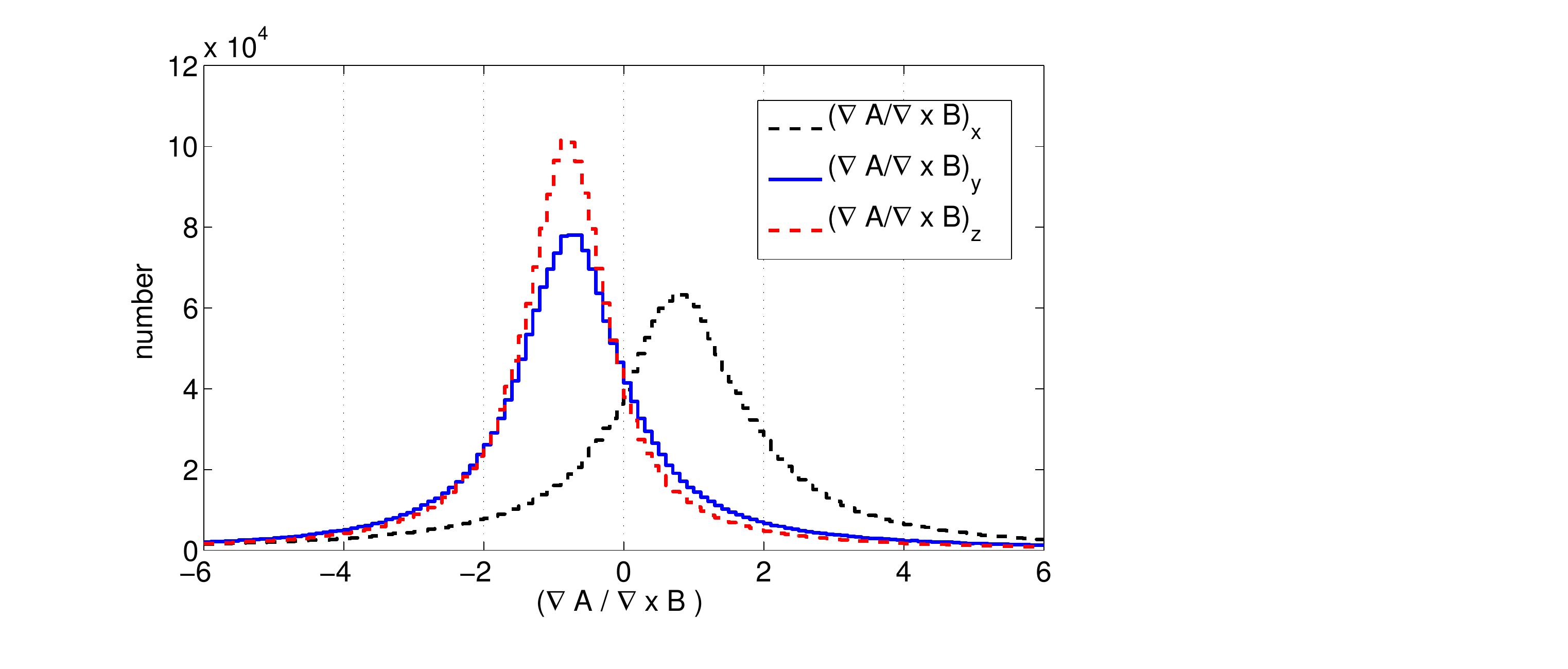}}
 \caption{A histogram of the ratios of irrotational to solenoid components of $({\bf \Psi} \cdot {\bf \hat{x}}){\bf \hat{x}}$ measured in our N-body simulation. The black line is the ratio projected along the ${\bf \hat{x}} $ axis with mean value of 0.75 and standard deviation, $\sigma=2.50$. The blue line shows the values projected along ${\bf \hat{y}}$ with mean of -0.75 and $\sigma=2.25$, and the red line shows the same projected along ${\bf \hat{z}}$ with mean of -0.85 and $\sigma$ =1.95. The differences between the blue and red curve are due to noise {\correction as only one simulation is used in this test.}}
 \label{fig:helmholtzHIST}
\end{figure}

To empirically measure the amplitude of these components we compute the three displacement field matrices that comprise ${\bf \Psi}_{\textrm{true}}$ at each grid point in the N-body catalogue. We smooth these using a Gaussian filter with smoothing scale $R=10\mpcoh$.\footnote{This smoothing length chosen is shown to perform best when reconstructing the linear BAO signal extracted from spherically averaged power spectrum measurements \citep{Burden14}.} We then compute the soleniodal and irrotational components of $({\bf \Psi\cdot \hat{x}} ){\bf \hat{x}}$ in Fourier space using FFTs as in Eq's.~\ref{eq:solenoidal}~and~\ref{eq:irrotational}. The measurements are converted back to configuration space for comparison.
Fig.~\ref{fig:Helmholtz_hist} shows scatter plots comparing the amplitudes of the vector fields in cartesian directions. A histogram of the ratios of irrotational to solenoid components is shown in Fig.~\ref{fig:helmholtzHIST}. 

As expected from Eq's.~\ref{eq:solenoidal}~and~\ref{eq:irrotational}, $(\nabla A) \cdot {\bf \hat{y}}$ and $(\nabla \times {\bf B}) \cdot {\bf \hat{y}}$ are negatively correlated and $(\nabla A) \cdot {\bf \hat{z}}$ and $(\nabla \times {\bf B}) \cdot {\bf\hat{z}}$ are negatively correlated as the RSD only lie along the x-axis. The amplitude of these terms is small compared with that along the x-axis.

The Helmholtz components projected along ${\bf \hat{x}}$ are positively correlated and have a larger amplitude. 
In the plane parallel approximation the irrotational and solenoidal components projected along ${\bf \hat{x}}$ have a mean ratio of 0.75. From this empirical result we make a new approximation that $\nabla A \cdot {\bf \hat{x}} \approx 0.75 (\nabla \times {\bf B})\cdot {\bf \hat{x}}$. 
With this approximation, assuming that the geometry of the survey allows the plane parallel approximation, we have that $\nabla A \approx (3/7)( {\bf \Psi }\cdot {\bf \hat{r}}) {\bf \hat{r}}$ so that Eq.~\ref{eq:approxpsi} becomes
\begin{equation}
 \mathbf{\Psi} + \frac{3f}{7} \left( \mathbf{\Psi \cdot \hat{r}} \right) \mathbf{\hat{r}}=-\nabla \nabla^{-2} \frac{\delta_g}{b},
\end{equation}
rather than Eq.~\ref{eq:approxpsi}.
This has the solution
\begin{equation}\label{eq:approx2}
{\bf \Psi } = {\rm IFFT} \left[- \frac{i\mathbf{k }\delta_g\left(\mathbf{k}\right)}{k^2 b}\right] - \frac{3f}{7+3f} \left\{ {\rm IFFT} \left[- \frac{i\mathbf{k }\delta_g\left(\mathbf{k}\right)}{k^2 b}\right] \cdot{\bf \hat{r}} \right\} {\bf \hat{r}}.
\end{equation}
This suggests that our initial approximation in Eq.~\ref{eq:approx1} over compensates for the RSD component of the displacement field. {\correction This approximation is derived empirically from one N-body realisation. To use this method for data analysis, tests are required over a large number of realisations with varying cosmological parameters.}
In Section~\ref{sec:iterative} we show that the approximations are not important for the end result if we adopt an iterative approach. They only affect speed of convergence.

\subsection{More general geometries}

 \begin{table}
 \centering
  \begin{tabular}{llllllllll}
   \hline
${\bf  \Psi}$ &  $ \left\langle\dfrac{{\bf \Psi}_{\textrm{real, FFT}} \cdot {\bf \hat{r}}}{ {\bf \Psi} \cdot {\bf \hat{r}}}\right\rangle $ & $ \sigma$& outliers [\%]\\ 
\hline  
\textcolor{blue}{N-body}\\
 ${\bf \Psi}_{\textrm{red, FFT}}$&0.79 &   0.28 &   2 \\
 ${\bf \Psi}_{\textrm{FFTA}}$(Eq.~\ref{eq:approx1})& 1.13 &   0.34&   3\\
 ${\bf \Psi}_{\textrm{FFTA}}$(Eq.~\ref{eq:approx2})& 0.96 &   0.31 &  2 \\
\hline
\hline
\textcolor{blue}{Galaxy mock}\\
 ${\bf \Psi}_{\textrm{red, FFT}}$&0.80 &   0.42 &   4 \\
 ${\bf \Psi}_{\textrm{FFTA}}$(Eq.~\ref{eq:approx2})& 1.04 &   0.47 & 5   \\
${\bf \Psi}_{\mathrm{red, FD}}$ &   1.12  &  0.48 &  6\\
${\bf \Psi}_{\mathrm{real, FD}}$ &   0.97  &  0.52 &  7\\
\hline
${\bf  \Psi}$ &  $ \left\langle\dfrac{{\bf \Psi}_{\textrm{real, FFT}} \cdot {\bf \hat{r}_\perp}}{ {\bf \Psi} \cdot {\bf \hat{r}_\perp}}\right\rangle $ & $ \sigma$& outliers [\%] \\ 
\hline
  ${\bf \Psi}_{\mathrm{FFTA} } $(Eq.~\ref{eq:approx2})  &  1.05  & 0.53 & 2 \\                     
 \hline 
\end{tabular}
  \caption{Mean ratio and standard deviation of the the real space displacement field to the displacement field computed using different estimations (see subsection~\ref{sub:list} for notation) projected in the radial direction for both N-body simulations (top) and mock galaxies (centre). Outliers, defined as ratio values above 3 or less than -1 have been cut from the analysis. The average values of the mean ratio and standard deviation between the estimated displacement field and real space displacement field projected in the off-radial directions (${\bf \hat{r}_\perp}$) is shown in the bottom row. }
   \label{table:tab1}
\end{table}
We compare the real displacement field projected in the radial direction, $({\bf \Psi}_{\textrm{real, FFT}} \cdot {\bf \hat{r}} )$ to the displacements computed using the non-corrected field $({\bf \Psi}_{\textrm{red, FFT}} \cdot {\bf \hat{r}} )$.
Then we compare the real displacement field projected in the radial direction to the displacements computed using the approximations in Eqs.~\ref{eq:approx1}~and~\ref{eq:approx2} projected in the radial direction, $({\bf \Psi}_{\textrm{FFTA}}\cdot {\bf \hat{r}} )$, using the N-body catalogue. The comparisons are quantified by taking the mean ratio of these values at each grid point. The mean ratios and standard deviations are shown in the top part of Table~\ref{table:tab1}. {\correction To prevent true displacements close to zero skewing the distribution, we cut out outlying ratios that are above 3 or below -1. The percentage of outliers in each computation is shown in the far right column of the table.}

With no correction, the mean ratio of the amplitude of the radial component of the real displacement field to that of the redshift-space displacement field is 0.79 with a standard deviation of 0.28. The enhanced clustering in the radial direction in redshift-space leads to an overestimation of the amplitude compared to the real space displacement vectors.
Using the approximation in Eq.~\ref{eq:approx1} the mean ratio between the amplitude of the real space field projected in the radial direction and the corrected displacement field is 1.13 with a standard deviation of 0.34. As predicted the displacement field in the radial direction is overcorrected. Using the approximation in Eq.~\ref{eq:approx2} the mean ratio between the amplitude of the real space field projected in the radial direction and the corrected displacement field is 0.96 with a standard deviation of 0.31. The amplitudes of the displacements are on average 5\% larger than those of the real space displacement. The approximation in Eq.~\ref{eq:approx2} therefore brings the estimate of the displacement field closest to the real space displacement field even though the N-body simulation does not have a plane-parallel RSD geometry.

Next we use the mock galaxies to compare the real displacement field projected in the radial direction, $({\bf \Psi}_{\textrm{real, FFT}} \cdot {\bf \hat{r}} )$ to the non-corrected displacement field, $({\bf \Psi}_{\textrm{red, FFT}} \cdot {\bf \hat{r}} )$, and the displacement field computed using the approximation in Eq.~\ref{eq:approx2}, $({\bf \Psi}_{\textrm{FFTA}}\cdot {\bf \hat{r}} )$ projected in the radial direction. Following this we compute the displacement field using the finite difference method outlined in subsection~\ref{sub:FD}, from the redshift-space overdensity. In order to compare the results of both methods in real and redshift-space we also compute the real space displacement field using finite differences.

With no correction, the mean ratio of the amplitude of the radial component of the real displacement field to that of the redshift-space displacement field is 0.80 with a standard deviation of 0.42. The ratio shows that (as seen in the N-body simulation), the displacements computed from the redshift-space data are larger, on average, than the real space displacement vectors.
Using the approximation in Eq.~\ref{eq:approx2} the mean ratio between the amplitude of the real space field projected in the radial direction and the corrected displacement field is 1.04 with a standard deviation of 0.47.
The approximation brings the estimate of the displacement field closer to the real space displacement field. The amplitude of the radial displacements are smaller by 4\% than the real space values.

The mean ratio of the real space displacement fields computed using FFTs and finite differences projected in the radial direction\footnote{The field is isotropic but we chose the radial direction for consistency.} is 0.97 with a standard deviation of 0.52. Therefore there is only a small difference between methods applied to real space data as expected.
The mean ratio of the amplitude of the real space displacement field and the redshift-space displacement field both computed in configuration pace and projected in the radial direction is 1.12 with a standard deviation of 0.48. Therefore the finite difference method overcorrects the redshift-space displacements in the radial direction. {\correction The accuracy of finite difference scheme is dependent on the resolution of the grid used. Finer grid resolution would improve the accuracy of the computations, however, the procedure would be more computationally expensive.}
The average mean ratios and standard deviations are shown for all of these results in the second part of Table~\ref{table:tab1}.

Finally, to check that the approximation in Eq.~\ref{eq:approx2} does not induce distortions in the off-radial directions (${\bf \hat{r}_\perp}$), we compute the ratios $({\bf \Psi}_{\textrm{real, FFT}} \cdot {\bf \hat{r}_\perp} )$ to $({\bf \Psi}_{\textrm{FFTA}} \cdot {\bf \hat{r}_\perp})$. 
The mean ratio is 1.05 with a standard deviation of 0.53. The values are the same for both off-radial directions. We conclude that any distortions cast by the approximation in the non-radial directions are negligible. The values are tabulated at the bottom of Table~\ref{table:tab1}.

In this section we have shown that the approximation $({\bf \Psi} \cdot {\bf \hat{r}}) {\bf \hat{r}}$ is irrotational as adopted by \cite{Burden14} overcorrects for the RSD component of the displacement field. However it is closer to the real space displacement field than the non-corrected field.
Assuming a plane parallel approximation of RSDs we empirically derived the relationship between the irrotational and solenoidal components and used this to improve our estimate the real space displacement field.
Our updated approximation is shown in Eq.~\ref{eq:approx2}. 
We show that this approximation, computed using FFTs, is closer to the real space displacement field than one computed in configuration space which overcorrects the RSD component of the displacement field.
 \begin{figure*}
 \begin{tabular}{cc}
    \resizebox{0.5\textwidth}{!}{\includegraphics{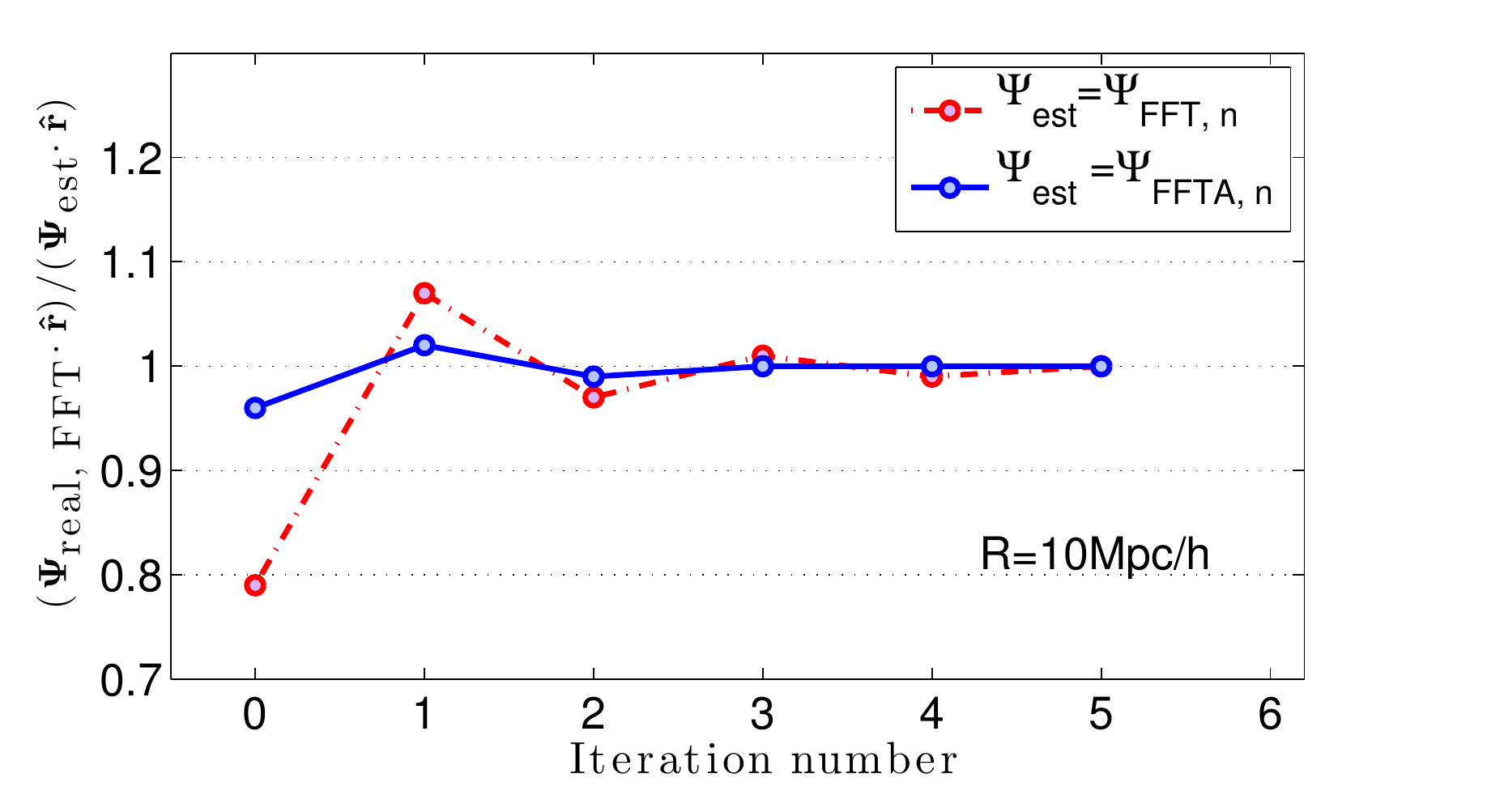}}&
    \resizebox{0.5\textwidth}{!}{\includegraphics{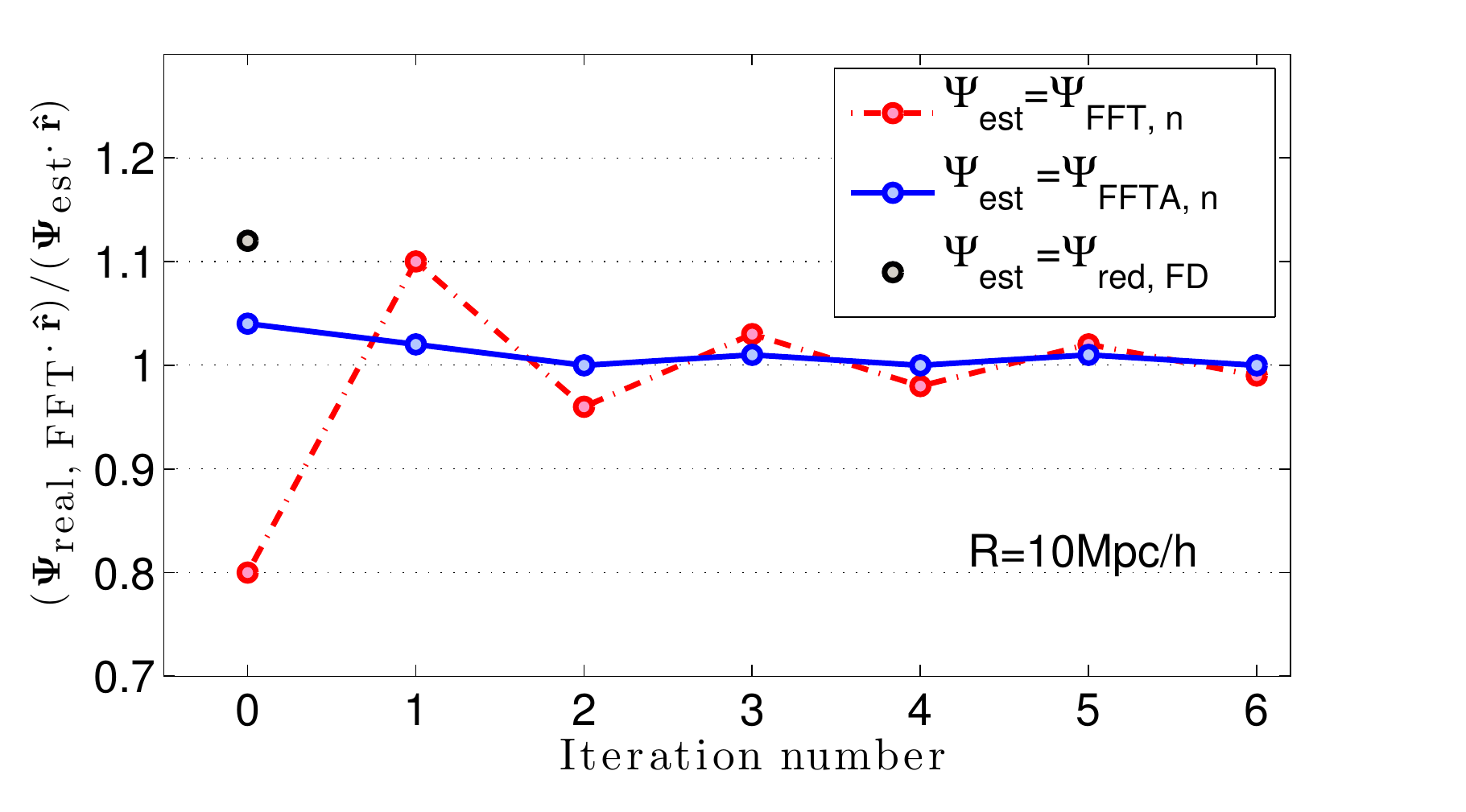}}\\
\end{tabular}
 \caption{Iterative convergence of the ratio of $({\bf \Psi}_{\textrm{real, FFT}}\cdot{\bf \hat{r}})$ to $({\bf \Psi}_{\mathrm{est}}\cdot{\bf \hat{r}})$ measured from the N-body simulation (left) and the mock galaxy catalogue (right). Starting with the non-corrected field (red dashed line), the N-body ratios at each iteration oscillate and converge after 4 iterations. Starting with the corrected field (blue full line), the values converge after 2 iterations. Starting with the non-corrected field (red dashed line), the mock catalogue ratio at each iteration oscillates about 1 and converges after 6 iterations. Starting with the corrected field (blue full line), the mock catalogue values converge at 2 iterations. {\correction The criterion to define convergence is stated in the main text.}The same ratio from the finite difference computation is 1.12 (grey point).}
 \label{fig:itmock}
\end{figure*}
\section{An iterative reconstruction scheme}\label{sec:iterative}

In this section we propose an iterative method of estimating the real space overdensity from the redshift-space overdensity field. {\correction Note that this iterative method is a precursor to the actual reconstruction process.} We apply the iterative method to both the N-body and mock catalogues. The ratio of the amplitude of the real space overdensity field to the amplitude of the displacement field computed at each iteration in the radial direction is compared. This iterative method allows us to avoid the approximations discussed in the previous section.

\subsection{Method}
As seen in Section~\ref{sec:effectsirrot}, we can estimate ${\bf \Psi}$ using Eq.~\ref{eq:approx2}, or we could have ignored the RSD completely and our estimate would be ${\bf \Psi} = \textrm{IFFT}\left[i{\bf k}\delta_{\textrm{g, red}}/k^2 b\right]$. In terms of the potential, our estimate is $\phi_{\textrm{est, 1}} = \textrm{IFFT}\left[ \delta_{\textrm{g, red}}/ k^2 b\right]$. From Eq.~\ref{eq:mainpot}, we expect our estimate to be related to the potential $\phi$ by
\begin{equation}
\phi_{\textrm{est, 1} }= \phi + fG(\phi).
\end{equation}
To simplify the algebra, we have defined an operator 
\begin{equation}
G(\phi) \equiv \nabla^{-2} \nabla \cdot ({\bf \nabla} \phi \cdot {\bf \hat{r}}){\bf \hat{r}}.
\end{equation}
The estimate of the potential can be used to remove the redshift component of the displacement field so that our first estimate of the real space overdensity is
\begin{equation}
\frac{\delta_{g, \textrm{real, 1}}}{b} = \frac{\delta_{g, \textrm{red}}}{b} + f\nabla \cdot ({\bf \nabla} \phi_{\textrm{est, 1}} \cdot {\bf \hat{r}}){\bf \hat{r}},
\end{equation}
which, substituting for $\phi_{\textrm{est, 1}}$ becomes
\begin{equation}
\frac{\delta_{g, \textrm{real, 1}}}{b} = -\nabla^2 \phi + f^2 \nabla \cdot (\nabla G(\phi)\cdot {\bf\hat{r}} ){\bf\hat{r}}.
\end{equation}
From this we can make a second estimate of the potential
\begin{equation}
\phi_{\textrm{est, 2}} = -\nabla^{-2}\frac{\delta_{\textrm{real, 1}}}{b} = \phi - f^2G(G(\phi)).
\end{equation}
As before, we use it and Eq.~\ref{eq:mainpot} to remove the redshift component of the displacement field giving us our second estimate of the real space overdensity
\begin{equation}
\frac{\delta_{g, \textrm{real, 2}}}{b} = \frac{\delta_{g, \textrm{red}}}{b} + f\nabla \cdot ({\bf \nabla} \phi_{\textrm{est, 2}} \cdot {\bf \hat{r}}){\bf \hat{r}}.
\end{equation}
Or equivalently
\begin{equation}
\frac{\delta_{g, \textrm{real, 2}}}{b} = -\nabla^2 \phi  - f^3\nabla \cdot ({\bf \nabla} G(G(\phi )) \cdot {\bf\hat{r}}) {\bf \hat{r}},
\end{equation}
and we see our recursive iterative method emerge. In general for each iteration n, the corrected overdensity is related to the true galaxy overdensity by
\begin{equation}\label{eq:iterative}
\frac{\delta_{g, \textrm{real, n}}}{b} = \frac{\delta_{\textrm{g, real}}}{b} + (- f)^{n+1} \nabla \cdot ({\bf \nabla} G_{(\textrm{n})}(\phi ) \cdot {\bf\hat{r}}) {\bf \hat{r}},
\end{equation}
where the subscript n in the $G_{(\textrm{n})}(\phi)$ term describes the number of times the function $G(\phi)$ is applied to itself.
We then go on to use $\delta_{g, \textrm{real, n}}$ to estimate ${\bf \Psi}_{\textrm{FFT, n}}= -\nabla \nabla^{-2}(\delta_{\textrm{g, real, n}}/b)$.

For a plane parallel approximation in Fourier space where ${\bf \hat{r}} \rightarrow {\bf \hat{x}}$ the n$^{\textrm{th}}$ iteration of the displacement field can be written as
\begin{equation}\label{eq:itpsi}
{\bf \Psi}_{\textrm{FFT, n}} = -\frac{i{\bf k} \delta_{\textrm{g, real}}}{b k^2}\left [ 1 + (-f)^{(\textrm{n} +1)}\left(\frac{k_x}{k}\right)^{2(\textrm{n}+1)}\right].
\end{equation}
For the iterative process to converge, the second term on the RHS of Eq.~\ref{eq:itpsi} must decrease in magnitude as n increases.
As $k_x \leq k$ it will be reduced in amplitude at each iteration provided $f<1$. Similar convergence is expected for more general geometries given that the relative amplitudes of the terms do not change.

\subsection{Results}
We consider two starting points for the iterative procedure. The first is the displacement field computed directly from the redshift-space coordinate using FFTs, ${\bf \Psi}_{\textrm{red, FFT}}$. {\correction The second is the same} field modified by {\correction the approximate correction given in Eq.~\ref{eq:approx2}}. We call this ${\bf \Psi}_{\textrm{FFTA}}$. The iterated versions of these fields are denoted ${\bf \Psi}_{\textrm{FFT, n}}$ and ${\bf \Psi}_{\textrm{FFTA, n}}$ respectively where n is the iteration number.
From Eq.~\ref{eq:iterative} we see the ratio of the displacement field measured from the real space overdensity projected in the radial direction to that of the displacement field computed from the redshift-space overdensity projected along the radial direction will be
\begin{equation}\label{eq:ratio}
\frac{{\bf \Psi}_{\textrm{real, FFT}}\cdot {\bf \hat{r}} } {{\bf \Psi}_{\mathrm{FFT, n}}\cdot {\bf \hat{r}}} =\frac{ {\bf \Psi}_{\textrm{real, FFT} }\cdot{\bf \hat{r}}} {  {\bf \Psi}_{\textrm{real, FFT}} \cdot{\bf \hat{r}} + (-f)^{(n+1)} {\bf G}_{\Psi}\cdot {\bf \hat{r}}},
\end{equation}
where, if the term 
\begin{equation}
{\bf G}_{\Psi} = -\nabla \nabla^{-2}\left[ \nabla \cdot ({\bf \nabla} G_{(n)}(\phi) \cdot {\bf\hat{r}}) {\bf \hat{r}}\right]
\end{equation}
 decreases in magnitude with each iteration, the ratio will oscillate around one and converge provided we use the correct growth value in the procedure. 
  \begin{table}
 \centering
  \begin{tabular}{llllllllll}
  \hline
${\bf  \Psi}_{\textrm{est}}$ &  n  & $ \left \langle\dfrac{{\bf \Psi}_{\textrm{real, FFT}} \cdot {\bf \hat{r}}}{ {\bf \Psi}_{\textrm{est}} \cdot {\bf \hat{r}}}\right \rangle $ & $ \sigma$& outliers [\%] \\ 
\hline                       
${\bf \Psi}_{\textrm{FFT, n}}$&0  &  0.79 &   0.28 &   2 \\
&1 &   1.07 &   0.50  &   7 \\
&2 &   0.97 &   0.32 &    3   \\
&3 &   1.01 &   0.40 &    4 \\
&4 &   0.99 &   0.36  &   3 \\
&5&    1.00 &   0.37  &   4 \\
 \hline
 ${\bf \Psi}_{\textrm{FFTA, n}}$&   0  &  0.96 &   0.31 &  2 \\
&    1 &   1.02      &   0.44 &  5\\
&    2 &    0.99     &   0.35&    3   \\
&    3 &   1.00      &   0.39 &   4\\
&    4 &   1.00      &  0.37&  4\\
&    5 &    1.00     &  0.37  &   4\\                        
 \hline 
\end{tabular}
  \caption{Iterative results for the N-body simulation. There are 20($128^3$) particles in the sample which has been placed on a $128^3$ grid and smoothed with R=10$\mpcoh$. The mean ratio and standard deviation of the ratio of the real space displacement field projected in the radial direction to that computed at each iteration are shown. The two starting points are the displacement field measured directly in redshift space, ${\bf \Psi}_{\textrm{FFT, (0)}}$ and the approximation in Eq.~\ref{eq:approx2}, ${\bf \Psi}_{\textrm{FFTA, (0)}}$. We define convergence as the iteration to which subsequent values of the mean differ by less than 1\%.
 Starting with no approximation the iterative procedure converges in 4 steps. Starting with the approximation the procedure converges in 2 steps. {\correction Outliers are defined as ratio values greater than 3 or less than -1.}}
 \label{table:iterative1}
\end{table}
\begin{table}
  \begin{tabular}{llllllllll}
  \hline
${\bf  \Psi}_{\textrm{est}}$ &  n  & $ \left \langle\dfrac{{\bf \Psi}_{\textrm{real, FFT}} \cdot {\bf \hat{r}}}{ {\bf \Psi}_{\textrm{est}} \cdot {\bf \hat{r}}}\right \rangle $ & $ \sigma$& outliers \\ 
\hline
${\bf \Psi}_{\textrm{FFT, n}}$ & 0  &  0.80 &   0.42  &   4\\
& 1 &   1.10 &   0.51&    7 \\
&  2 &   0.96 &   0.43 &  4  \\
&    3 &   1.03 &   0.47 &  6 \\
&  4 &   0.98 &   0.43  &  5 \\
&  5 &   1.02 &   0.46  &  5 \\
&  6 &    0.99  &  0.44&  5 \\
 \hline
${\bf \Psi}_{\textrm{FFTA, n}}$ &0  &  1.04 &   0.47 &   5 \\
&    1 &   1.02 &   0.46  &   5 \\
&    2 &   1.00 &   0.43&    5   \\
&    3&   1.01 &   0.45&    5   \\   
&    4 &   1.00 &   0.44 &  5   \\       
&    5 &   1.01 &   0.45 &   5  \\      
&    6 &   1.00 &   0.44 &   5  \\                    
 \hline                       
\end{tabular}
  \caption{Iterative results for the mock catalogue. There are 601921 galaxies in the sample which has been placed on a $512^3$ grid and smoothed with R=10$\mpcoh$. Shown is the mean ratio and standard deviation of the ratio of the real space displacement field projected in the radial direction to that computed at each iteration projected in the radial direction. The two starting points are the displacement field measured directly in redshift space, ${\bf \Psi}_{\textrm{FFT, (0)}}$ and the approximation in Eq.~\ref{eq:approx2}, ${\bf \Psi}_{\textrm{FFTA, (0)}}$. We define convergence as the iteration to which subsequent values of the mean differ by less than 1\%.
 Starting with no approximation the iterative procedure converges in 6 steps. Starting with the approximation the procedure converges in 2 steps. {\correction Outliers are defined as ratio values greater than 3 or less than -1.}}
    \label{table:iterative2} 
  \end{table}

In Fig.~\ref{fig:itmock} we show how the mean of the ratio in Eq.~\ref{eq:ratio} computed at each particle/galaxy position changes for successive iterations in both N-body and mock catalogues. For comparison we also show the ratio computed from the finite difference algorithm applied to the mock catalogue.
The values are provided in Table~\ref{table:iterative1} (N-body) and Table~\ref{table:iterative2} (mock).

As predicted, the mean values of the ratios $({\bf \Psi}_{\textrm{real, FFT}}\cdot {\bf \hat{r}})$ to $({\bf \Psi}_{\mathrm{FFT, n}}\cdot {\bf \hat{r}})$ oscillate around one and converge. Note that we compare smoothed fields and the smoothing is a convolution that does not cancel out in the ratio. We define convergence as the point at which further iterations do not change the mean value of the ratio by more than 1\%.
There is no further improvement after iteration number 4 for the N-body simulation and iteration number 6 for the mock data sets. 
In the N-body catalogues the mean ratio converges to 0.99 with a standard deviation of 0.36. In the mock galaxies the mean ratio converges to 0.99 with a standard deviation of 0.44. 
Starting the iteration with the corrected field, 2 iterations are needed for the ratio to converge to 0.99 with a standard deviation of 0.35 in the N-body simulation. Using the galaxy mock sample 2 iterations are needed for the mean ratio to converge to 1.00 with a standard deviation of 0.43. {\correction As in Table \ref{table:tab1} the column on the far right shows the percentage of outliers that have a ratio greater than 3 or less than -1 and have not been included in the computations.}

\begin{figure*}
    \resizebox{1.0\textwidth}{!}{\includegraphics{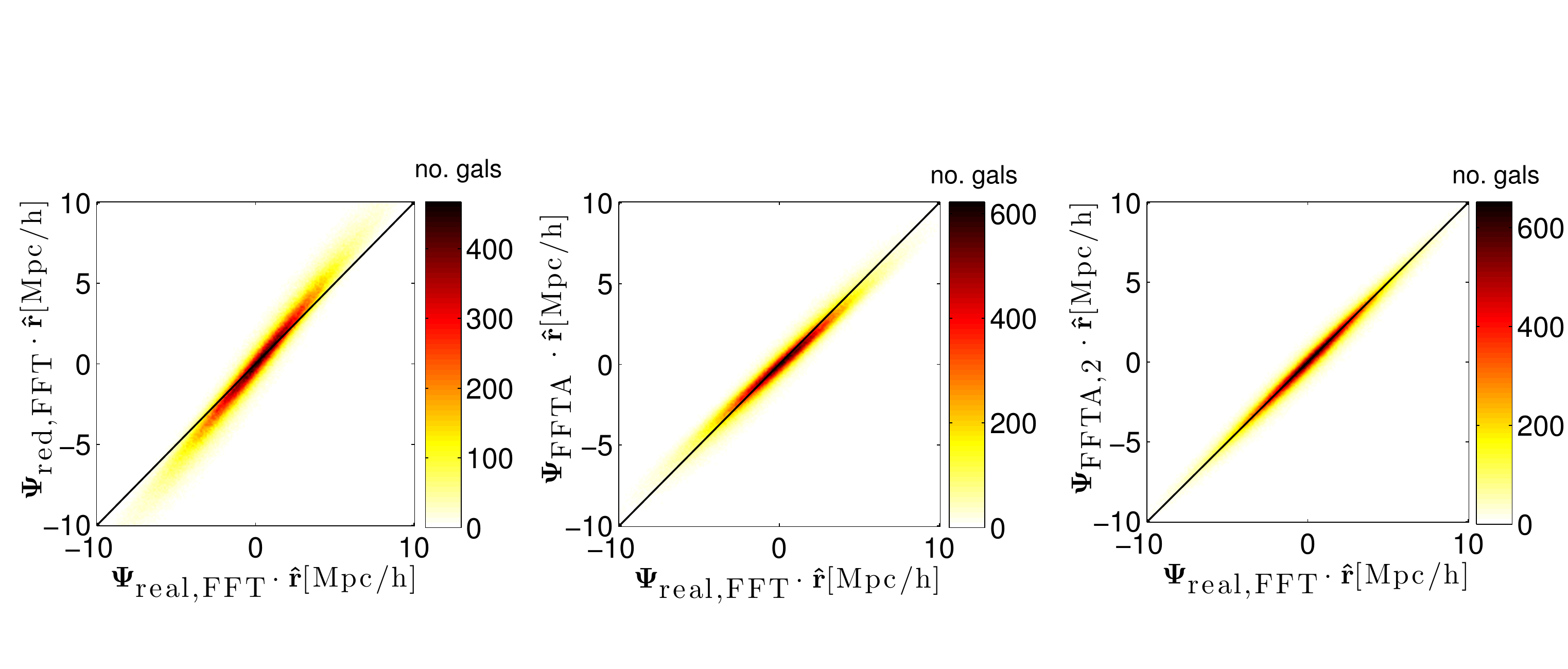}}
 \caption{From left to right the plots show a comparison of the real space displacement vectors projected in the radial direction to; the field computed directly from the redshift-space overdensity ${\bf \Psi}_{\textrm{red, FFT}}$; the field computed with the approximation given in Eq.~\ref{eq:approx2}, ${\bf \Psi}_{\textrm{FFTA}}$;  and the field computed by iterating twice starting with the same approximation, ${\bf \Psi}_{\textrm{FFTA,2}}$. The redshift-space field overestimates the displacement field, the approximation moves the two distributions closer together and the iterated field shows the tightest correlation with the real space field. \correction The 1:1 values are described by the black lines.}
 \label{fig:it1it4}
\end{figure*}

Fig.~\ref{fig:it1it4} compares the amplitude of ${\bf \Psi}_{\textrm{real, FFT}} \cdot {\bf \hat{r}}$ to ${\bf \Psi}_{\textrm{red, FFT}}\cdot{\bf \hat{r}}$, ${\bf \Psi}_{\textrm{FFTA}}\cdot{\bf \hat{r}}$ and ${\bf \Psi}_{\textrm{FFTA,2}}\cdot{\bf \hat{r}}$ computed from the mock catalogues at each galaxy position. It is seen that the tightest correction with the real space displacement field comes from the ${\bf \Psi}_{\textrm{FFTA,2}}\cdot{\bf \hat{r}}$ field.

\section{Lognormal transform of $\delta$}\label{sec:lognormal}
\begin{figure*}
    \resizebox{2.2\columnwidth}{!}{\includegraphics{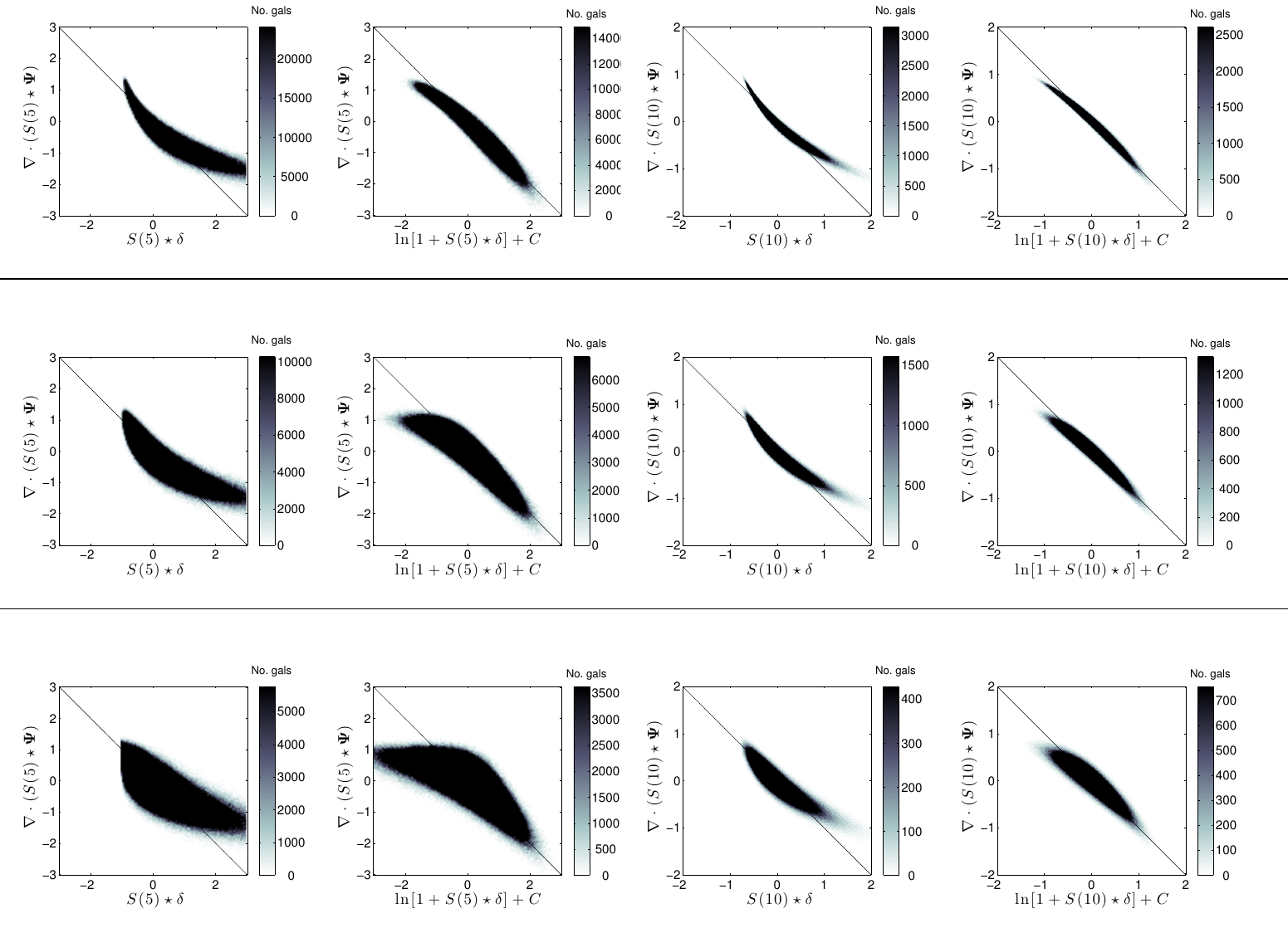}} 
   \caption{Scatter plots of the amplitude of the divergence of the Lagrangian displacement field at each grid point versus the $\delta$ field and the logarithmic transform fields. Different density samples are separated by the horizontal lines, the samples are top to bottom, 20t, 1t and 0p25t. The two plots on the left show the divergence of the Lagrangian displacement field at each grid point versus the the $\delta$ (far left) and versus the the $\delta$ field and the logarithmic transform (mid left) for smoothing $R=5\mpcoh$. The plots on the right show the same but for smoothing length $R=10\mpcoh$. 
 It is seen that the lognormal transform is better matched to the divergence field in the higher density samples. However, in the 0p25t sample when the divergence field is positive, the lognormal transform gets pulled to higher amplitude negative values and performs badly. This is more apparent at smaller smoothing scales as expected. Therefore the lognormal transform only improves our estimate of ${\nabla \cdot \bf\Psi}$ in high density regions.}
   \label{hist2dlognormal}
\end{figure*}

It has been shown \citep{Falck12} that under certain conditions, a logarithmic transform of the overdensity provides a better estimate of the divergence of the displacement field in N-body simulations. {\correction In \citep{Falck12}, the smoothing of the overdensity field is achieved using a CIC binning assignment, however we use a Gaussian smoothing kernel in this work, the goal of which is to explore the effect of sample density of the lognormal results. It should be noted that our densest sample is approximately 20 times less dense than the simulations used in their work.}
In this section we consider if such a transform can help with our reconstruction algorithm using different density N-body samples. 
We first test the transform in the calculation of $\nabla \cdot {\bf \Psi}$. We then test if the lognormal transform combined with the Zel'dovich approximation improves our estimate of ${\bf \Psi}$.
All of the comparisons are performed in real space. 
\begin{table}
\centering
  \begin{tabular}{llllllllll}
  \hline
Sample&  R[$\mpcoh$]  & $ \sigma_A$ &  $ \sigma_B$ \\ 
\hline                       
20t  & 10 &0.57  &0.43  \\
5t  & 10 &0.61  &0.49  \\
1t & 10 &0.79  & 0.70 \\
0p25t  & 10 &1.27  &1.21  \\
0p065t &10 &2.26 &2.48 \\
20t  & 5 &1.01  & 0.74 \\
5t  & 5 &  1.05& 0.82 \\
1t  & 5 & 1.28 & 1.16 \\
0p25t  & 5& 1.89  &2.15  \\
0p065t & 5&3.28&5.66\\
 \hline
 \end{tabular}
  \caption{ The standard deviation computed form the distributions ${\bf \Psi}_{\textrm{true}} -{\bf \Psi}(\delta)$ denoted $\sigma_A$ and $ {\bf \Psi}_{\textrm{true}} -{\bf \Psi}(\ln[1+\delta]+C)$ denoted $\sigma_B$.  As the density of the sample is decreased, the improvement seen by using the the lognormal transform of the overdensity to compute the displacement field decreases. {\correction For both the $5\mpcoh$ and $10\mpcoh$ smoothing scale at the lowest density sampled, the lognormal transform performs worse than the overdensity when used in the Zel'dovich approximation.}}
  \label{table:psi}

  \end{table}

The Newtonian continuity equation relates the rate of change of a density perturbation to its peculiar velocity ${\bf v}= a\dot{{\bf x}}$. It can be expressed (retaining the complete $(1+\delta)$ term in the time derivative \citep{Peebles80} )as
\begin{equation}
\frac{\partial(1+ \delta)}{\partial t} + \frac{1}{a}\nabla \cdot (1 + \delta) {\bf v} =0.
\end{equation}
This equation can be linearised about the quantity $\delta$ or $(1 + \delta)$. The respective linear continuity equations read
\begin{equation}
\frac{\partial \delta}{\partial t} + \frac{1}{a} \nabla\cdot{\bf v} =0,
\end{equation}
\begin{equation}
\frac{1}{(1+\delta)}\frac{\partial (1+ \delta)}{\partial t} + \frac{1}{a}\nabla\cdot {\bf v} =0,
\end{equation}
and have solutions
\begin{equation} \label{eq:delta}
\nabla \cdot {\bf \Psi} ^{(1)} = -\delta^{(1)},
\end{equation}
\begin{equation}\label{eq:logdelta}
\nabla \cdot {\bf \Psi }^{(1)} = -\ln (1 + \delta^{(1)}) + C.
\end{equation}

For each N-body density subsample, the overdensity is computed and smoothed with $R=5\mpcoh$ and $R=10\mpcoh$. 
The lognormal transform of the overdensity field is computed, $\ln ( 1 + S \star \delta) + \textrm{C}$ using the smoothed $\delta$ values. 
Following \cite{Falck12} we choose $\textrm{C}= - \langle \ln ( 1 + S \star \delta) \rangle$.
The displacements, ${\bf \Psi}_{\textrm{true}}$ computed using the 20t sample are smoothed with the same values of $R$. The divergence of the displacement field is computed in Fourier space, $\nabla \cdot (S \star {\bf \Psi}_{\textrm{true}})$. We compare it to our estimates of the displacement field. 
As one cannot compute the lognormal transform in void regions where $\delta=-1$ we use the smoothed density field rather than smoothing the whole term. This ensures that $\delta$ {\correction although converges to} never reaches -1.

\begin{figure*}
    \resizebox{2.0\columnwidth}{!}{\includegraphics{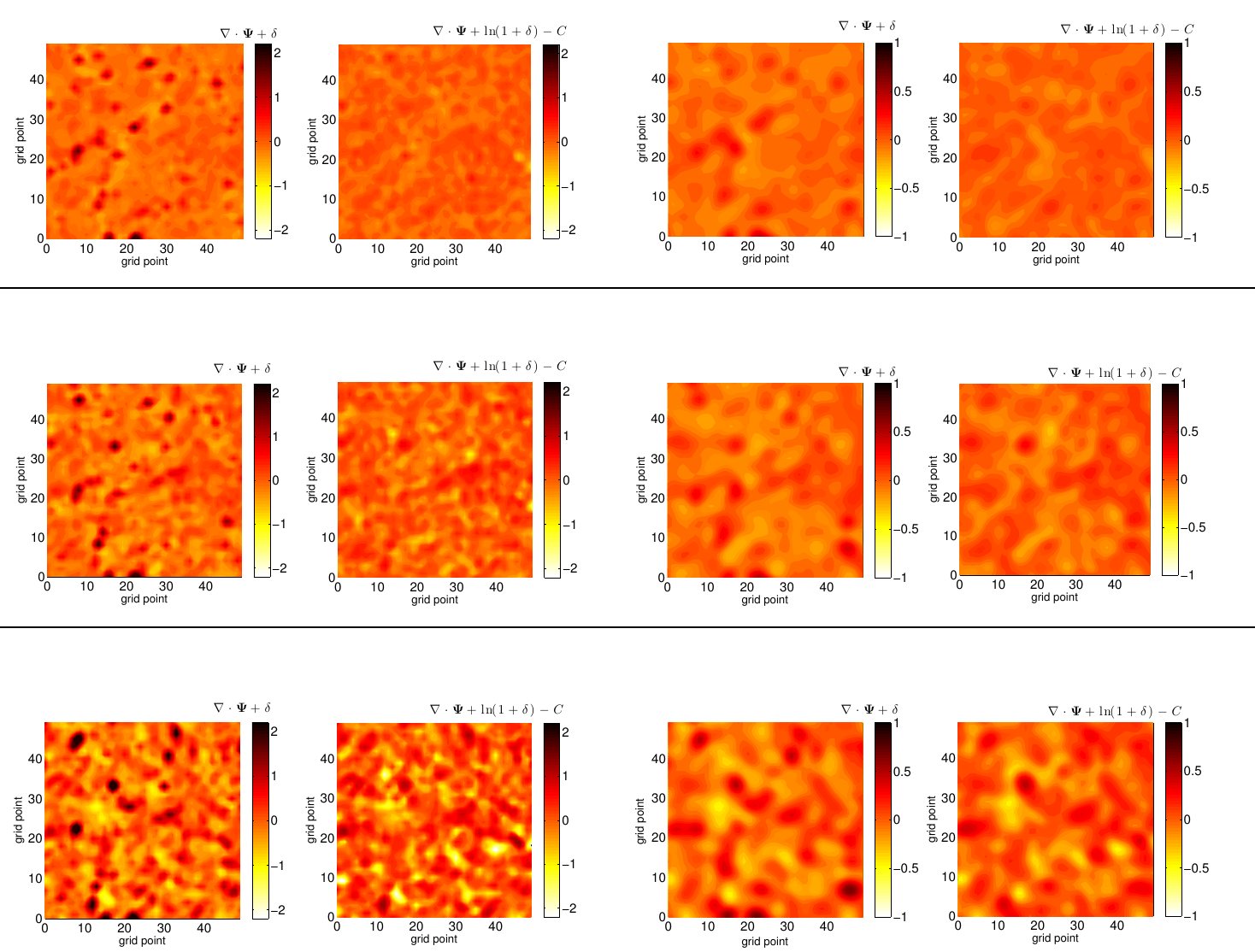}}
   \caption{The plots show a $1 \times 50^2$ grid point slice through the box for the divergence field measured from the 20t density sample plus the overdensity field or the lognormal transform of the overdensity for different density samples and smoothing lengths. The best match between fields will have the flattest contours and be closest to zero. 
The 20t sample (top) 1t sample (middle) and the 0p25t sample (bottom) are separated by horizontal lines. The column on the left shows the values smoothed $R=5\mpcoh$, and on the right smoothed with $R=10\mpcoh$.
 It is seen that for the higher densities, the lognormal transform clearly outperforms the overdensity at both smoothing lengths. {\correction This can be seen by comparing the hot-spots that emerge in the higher density regions within each sample.} As sample density is reduced the lognormal transform deviates from the divergence of the displacement field more than the overdensity in under dense regions. {\correction This can be seen as the blue cold-spots that emerge in the sparse regions of the samples.}The lognormal transform continues to provide a better match to the divergence of the displacement field in dense regions in the same sample. Reducing the smoothing scale increases the differences between the in the two field as expected.}
    \label{fig:contour} 
\end{figure*}
 
Scatter plots of the values of ${\bf \nabla} \cdot (S \star {\bf \Psi}_{\textrm{true}})$ versus the overdensity and logarithmic transform of the overdensity are shown in in Fig.~\ref{hist2dlognormal}.
In the higher density samples the lognormal transform is more correlated with the divergence of the true displacement field than the $\delta$ field alone. High density regions push the $\delta$ values above the values of $\nabla\cdot (S\star {\bf \Psi_\textrm{true}})$. The lognormal transform suppresses this effect. 
However in void regions where the smoothed overdensity tends towards -1 (more so for smaller smoothing scales), the lognormal transform varies rapidly to large negative values. 

Fig.~\ref{fig:contour} shows a $1 \times 50^2$ grid point slice through the N-body box. The contours show the difference in amplitude between ${\bf \nabla} \cdot (S \star {\bf \Psi}_{\textrm{true}})$ and both density field estimates for three of the samples at both smoothing lengths. The improvement from the lognormal transform of $\delta$ compared to $\delta$ at estimating the divergence of the true displacement field is reduced with sample density.

In summary, the lognormal transform straightens the relationship between density and the divergence of the displacement field for dense samples as previously shown in \cite{Falck12}. However, the method suffers from discreteness issues and provides a worse match to the divergence of the true displacement field in low density regions compared to the overdensity. {\correction These test are only carried out on the real space distributions and RSDs are ignored.}

In the second part of this section we test if the lognormal transform of the overdensity provides a better estimate of the displacement field, computed using the Zel'dovich approximation, than the overdensity.
We compute the displacement fields from Eq.~\ref{eq:Zeldovich}, where both the smoothed overdensity and the lognormal transform of the smoothed overdensity are used as a proxy for the linear overdensity field.
We compare these fields directly to $S\star {\bf \Psi}_{\textrm{true}}$. 

\begin{figure*}
 \resizebox{2.2\columnwidth}{!}{\includegraphics{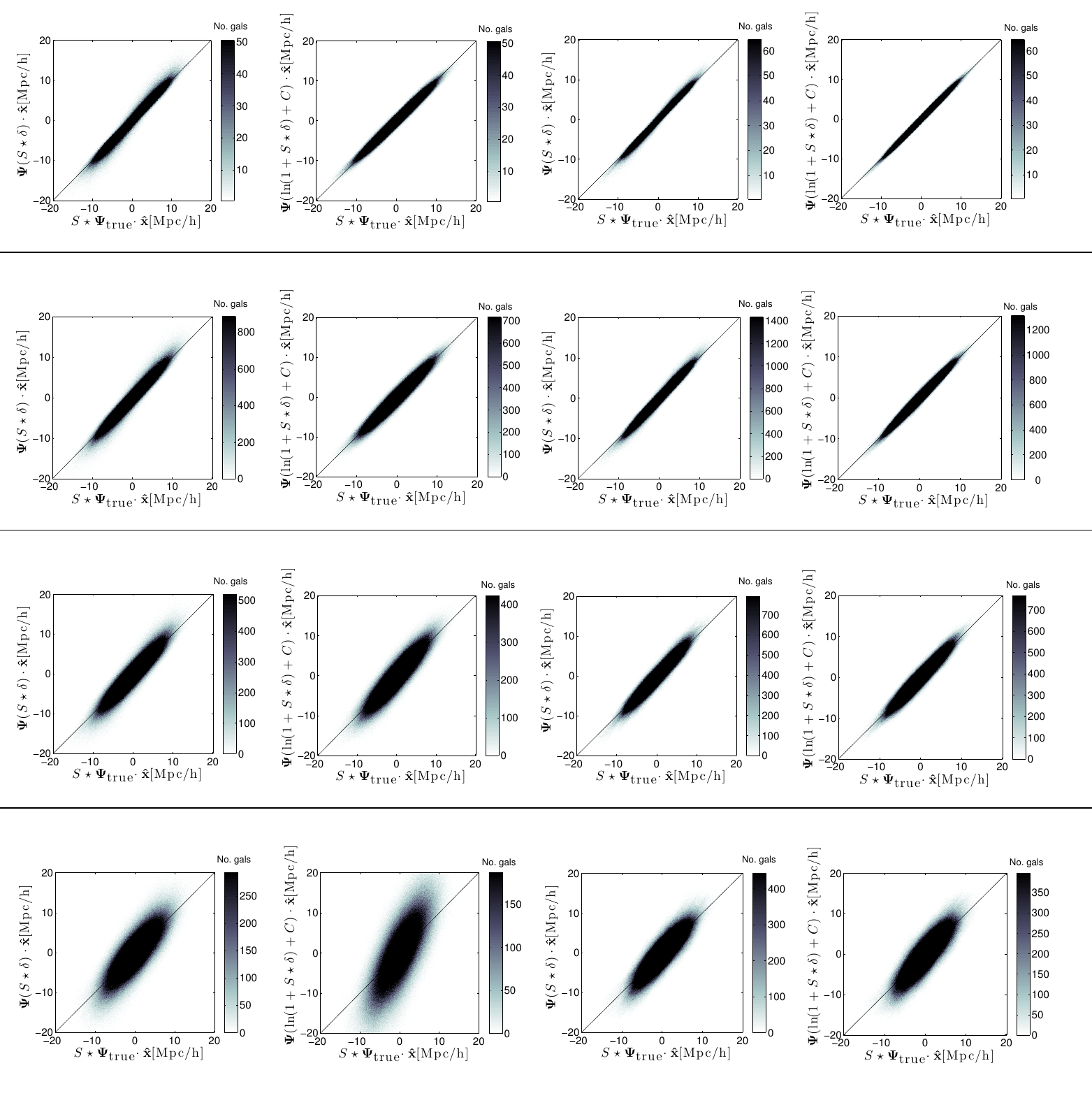}}
   \caption{Scatter plots of the smoothed Lagrangian displacement field at each grid point versus the displacements computed from the smoothed $\delta$ field and the logarithmic transform of smoothed $\delta$. Different density samples are separated by horizontal lines. The samples from top to bottom are 20t, 1t, 0p25t and 0p065t. The left two plots are smoothed with $R=5\mpcoh$ and are the displacements computed from $\delta$ and the lognormal transform of $\delta$ respectively. The plots on the right show the same but for smoothing length $R=10\mpcoh$. The displacement field computed from the lognormal transform of $\delta$ has a tighter correlation with true displacement vectors in the high density samples. {\correction However, in the 0p25t sample the displacements calculated from the lognormal transform of the overdensity smoothed with $R=5\mpcoh$ and in the 0p065t sample the displacements calculated from the lognormal transform of the overdensity smoothed with $R=5\mpcoh$ and $R=10\mpcoh$ do not match the displacement vectors as well as those computed from the smoothed overdensity.}}
   \label{fig:histpsi}
\end{figure*}

 Fig.~\ref{fig:histpsi} shows contour scatter plots of $S\star {\bf \Psi}_{\textrm{true}} \cdot {\bf \hat{x}}$ compared to both ${\bf \Psi}_{\textrm{true}}[S\star\delta({\bf x})]\cdot {\bf \hat{x}}$ and  ${\bf \Psi}_{\textrm{true}} [\ln (S\star \delta({\bf x}) + 1) +C]\cdot {\bf \hat{x}}$ for a subsampled range of smoothing lengths and sample densities. We create a new catalogue from the N-body simulations that has a realistic galaxy survey average number density $\bar{n}$, where $\bar{n} \approx 3 \times 10^{-4} h^3{\rm Mpc}^{-3}$. This catalogue has 0.065(128$^3$) particles and is called the 0p065t sample.
 We chose the ${\bf \hat{x}}$ direction as the distributions are in real space thus isotropic. 
 We quantify our results by computing the standard deviations of the differences between these distributions averaged over all cartesian directions, the values are shown in Table~\ref{table:psi}.
 The displacements computed from the smoothed overdensity in the sample with a realistic galaxy survey density, 0p065t, smoothed with a typical smoothing length for reconstruction, $R=10\mpcoh$, are better matched to the true displacements than those computed from the lognormal transform of the smoothed overdensity. 

\begin{figure}
 \resizebox{0.9\columnwidth}{!}{\includegraphics{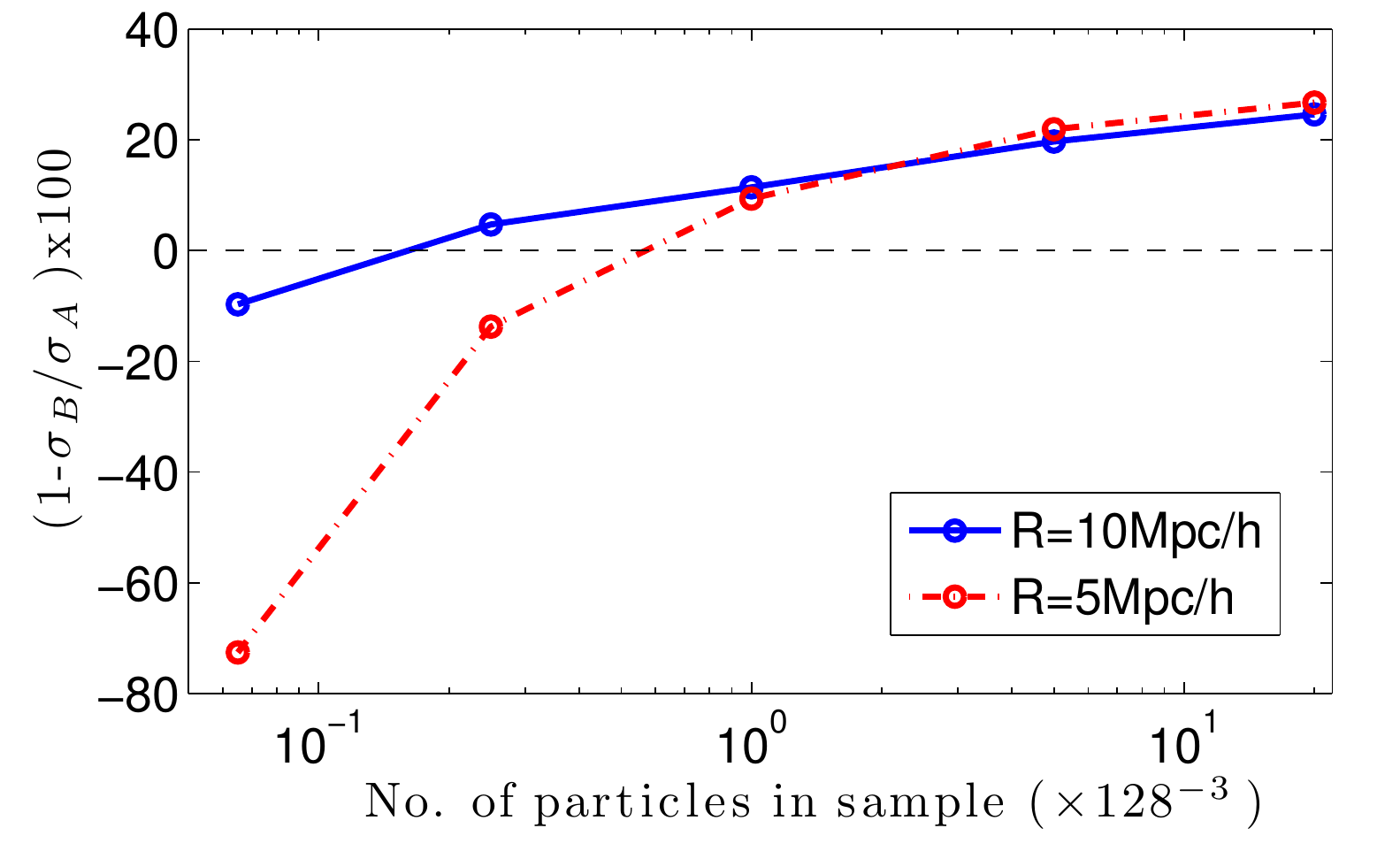}}
   \caption{The improvement in the estimate of the displacement field from using a lognormal transform of the overdensity is represented by $(1-\sigma_B/\sigma_A)\times 100$ where $\sigma_A$ is the standard deviation of the values of ${\bf \Psi}_{\textrm{true}}-{\bf \Psi}(\delta)$ and $\sigma_B$ is the standard deviation of the values of ${\bf \Psi}_{\textrm{true}}-{\bf \Psi}(\ln (1+ \delta) +C )$. The improvement decreases rapidly with decreasing sample density, $\bar{n}$. At the lowest density, $\bar{n}\approx3\times^{-4}\mpcoh^{-3}$, where there are $0.065(128^3)$ particles in the sample, the displacements computed from the logarithmic transform of the overdensity smoothed with both $R=10\mpcoh$ and $R=5\mpcoh$ do not match the smoothed true displacements as well as those computed from the smoothed overdensity. }
   \label{fig:sigma_density}
\end{figure}

In Fig.~\ref{fig:sigma_density} we show the improvement in the estimation of ${\bf \Psi}_{\textrm{true}}$ from a logarithmic transform of the overdensity compared to using the overdensity as a function of sample density. 
The values plotted are 1-$\sigma_B/\sigma_A$, where $\sigma_A$ is the standard deviation of the values of ${\bf \Psi}_{\textrm{true}}-{\bf \Psi}(\delta)$ and $\sigma_B$ is the standard deviation of the values of ${\bf \Psi}_{\textrm{true}}-{\bf \Psi}(\ln (1+ \delta) +C )$ computed at each grid point. The improvement rapidly drops off as the sample becomes more sparse. At the lowest density in the 0p065t catalogue, the displacements computed from the logarithmic transform of the overdensity smoothed with both $R=5\mpcoh$ and $R=10\mpcoh$ do not match the true smoothed displacements as well as those computed from the smoothed overdensity. The decline in improvement is steeper for smaller smoothing scales as the sample density is reduced.
As realistic galaxy surveys have densities of the same magnitude as the 0p065t sample, we conclude that it is not beneficial to use the logarithmic transform of the overdensity to compute the displacement vectors for reconstruction.

\section{Conclusion} \label{sec:conclusion}
Reconstruction is an important part of the analysis procedure in galaxy surveys designed to detect the BAO feature. Although it is shown to be a robust technique \citep{Burden14}, the standard algorithms must be improved for use in future galaxy survey data analysis.

In this paper we have looked at techniques that remove the redshift component of the displacement field computed from the Zel'dovich approximation applied to the overdensity field measured in redshift-space. We have also tested if a lognormal transform of the overdensity in the Zel'dovich approximation brings the values of the displacement field closer to those of the true Lagrangian displacement vectors.
{\correction It should be taken into consideration that tests were carried out on a single N-body and mock galaxy catalogue realisation assuming a single cosmology. Furthermore the mock galaxy catalogue was created using 2LPT, thus one should expect the 1LPT displacement vectors to be recovered to higher accuracy than a sample including the full non-linear gravitational evolution.}
We summarise the finding of the previous sections;
\begin{itemize}
\item The approximation that the redshift component of the displacement field is irrotational allows one to compute the displacement field using FFTs. It is better than no RSD correction. However, it is shown to overcorrect the RSD slightly. Previous work has shown that this does not affect the spherically averaged power spectrum measurements \citep{Burden14}. However it is not clear how reconstruction affects the anisotropic statistics. It is possible that the overcorrection of RSD due to this approximation may affect quadrupole measurements.
\item Empirical computations of the Helmholtz components of a plane parallel RSD component to the displacement field provide us with an improved approximation. The new approximation accounts for the irrotational components and an estimate of the amplitude of the solenoidal components of the RSD term. It is shown to be a better match to the real space displacement field in both our N-body simulation and mock galaxy catalogue.
\item We propose a fast iterative scheme to recover the real space displacement field (provisional on knowing the growth rate). Proof of convergence is shown for a plane-parallel RSD geometry in Fourier space and we expect the behaviour to hold for non-plane parallel geometries. Our iterative method does not introduce any distortion in the off-radial components in the recovered displacement field as we always start from our original overdensity and remove RSD using our estimate of ${\bf\Psi}$ in the radial direction only.
The scheme converges in two iterations for the mock sample and N-body catalogue when our starting point is the improved approximation above. The iterative algorithm does not require the approximations discussed above, however they speed up convergence.
\item The lognormal transform of the overdenisty helps to recover the true displacement field in high density simulations. However is does not perform well when computing the displacement field at realistic galaxy survey densities.
\end{itemize}
{\correction In these tests, the new iterative Fourier algorithm out-performs the configuration space method of recovering the real space displacement field as the accuracy of the configuration space method is limited by the resolution of the grid on which the overdensity is computed. However, the configuration space method performs very well in terms of the reconstruction process and it has not yet been shown that the new Fourier algorithm will offer any significant improvement in the effectiveness of the algorithm for current data.}
The effect of the new reconstruction algorithm on the anisotropic distribution of the overdenisty has not yet been fully analysed. It will be interesting to see if the methods described in this paper affect the post-reconstruction quadrupole in different ways. We suggest a list of questions to be addressed in future work.  
\begin{itemize}
\item How well does the iterative FFT reconstruction scheme improve the spherically averaged BAO measurement?
\item How do the methods outlined affect the anisotropic BAO measurements?
\item Can we use reconstruction to improve RSD measurements?
\end{itemize}
{\correction As galaxy positions are inferred from their redshift measurements, the background Hubble flow and the peculiar motion of a galaxy due to gravity are coupled along the line of sight. This distorts the galaxy clustering map enhancing the clustering signal on large scales along the line of sight. The anisotropic clustering signal therefore contains information about the growth rate of structure. It is possible that the reconstruction technique can be manipulated and applied to the the RSD field to increase the precision of these measurements.}
Despite these questions, we have shown that our new FFT reconstruction algorithm allows for higher resolution, more accurate measurements of the displacement field from the redshift-space galaxy overdensity than algorithms currently in use. This method will be very useful in the analysis of data expected from the plethora of upcoming galaxy surveys. 

\section*{Acknowledgments}
Many thanks to Patrick McDonald, Nikhil Padmanabhan, Mark Neyrink and Bridget Falck for useful correspondence. {\correction Thank you to the anonymous referee who provided a very helpful and comprehensive referee report.}
AB thanks the UKSA for support to develop algorithms for the Euclid Mission. WJP acknowledges support from the UK STFC through the consolidated grant ST/K0090X/1, and from the European Research Council through grant Darksurvey. CH is grateful for funding from the UK STFC. 
\bibliography{Fourier_recon_v1}

\end{document}